\documentclass[prl,twocolumn,superscriptaddress,showpacs,aps]{revtex4-1}
\pdfoutput=1
\usepackage{graphicx}

\begin{document}
\title{Single-Photon Absorption in Coupled Atom-Cavity Systems}
\author{Jerome Dilley}
\affiliation {University of Oxford, Clarendon Laboratory, Parks Road, Oxford, OX1 3PU, UK}
\author{Peter Nisbet-Jones}
\affiliation {University of Oxford, Clarendon Laboratory, Parks Road, Oxford, OX1 3PU, UK}
\author{Bruce W. Shore}
\affiliation {University of Oxford, Clarendon Laboratory, Parks Road, Oxford, OX1 3PU, UK}
\affiliation{Permanent address: 618 Escondido Circle, Livermore CA 94550, USA}
\author{Axel Kuhn}
\email{axel.kuhn@physics.ox.ac.uk}
\affiliation {University of Oxford, Clarendon Laboratory, Parks Road, Oxford, OX1 3PU, UK}
\date{\today}
\begin{abstract}
We show how to capture a single photon of arbitrary temporal shape with one atom coupled to an optical cavity. Our model applies to Raman transitions in three-level atoms with one branch of the transition controlled by a (classical) laser pulse, and the other coupled to the cavity. Photons impinging on the cavity normally exhibit partial reflection, transmission, and/or absorption by the atom. Only a control pulse of suitable temporal shape ensures impedance matching throughout the pulse, which is necessary for complete state mapping from photon to atom.  For most possible photon shapes, we derive an unambiguous analytic expression for the shape of this control pulse, and we discuss how this relates to a quantum memory.

\end{abstract}
\maketitle

\begin{figure}[h]
\centering\includegraphics[width=0.95\columnwidth]{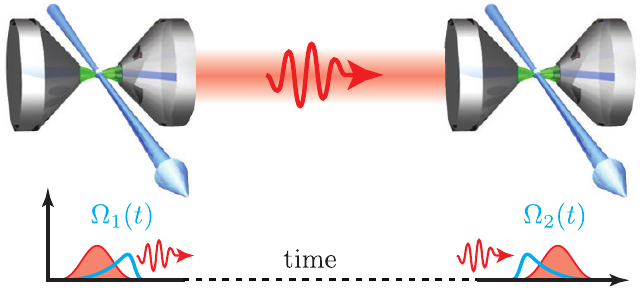}
\caption{(Color online) The Quantum Internet, showing the state of one atom (stationary qubit) being transferred via a photon (flying qubit) to that of another atom in a separate cavity.}
\label{qinet}
\end{figure}

In their seminal paper on the quantum internet, J.\,I.\,Cirac et al.\,\cite{Cirac97} introduced a scheme for photon-mediated state mapping between two distant atoms placed in spatially separated optical cavities. The experimental attainment of this would have important consequences for the fields of quantum information processing (QIP) \cite{DiVincenzo98,Knill01}, quantum networking, and the ultimate realisation of a ``quantum internet'' \cite{Kimble08}, where a single network link involves the state mapping from atom to photon in a first cavity, and then back to another atom in a second cavity, as illustrated in fig.\,\ref{qinet}. So far, good progress has been made experimentally: entanglement mapping mediated by photons with atomic ensembles \cite{Choi08,Matsukevich06}, cavity QED atom-photon state mapping  \cite{Wilk07-Science,Maitre97,BBMEJ07}, single-atom single-photon absorption in ion traps \cite{Piro10} and single-photon absorption by a coupled atom-cavity system \cite{Mucke10, Specht11} have all been demonstrated. However, no one has yet managed to combine all of the elements necessary to implement the scheme in its entirety. Further, there exists no general method for determining the optimal control of the classical laser pulses used in the generation and absorption of single photons in these processes, although some work has been done towards this \cite{Keller04,Santori10,Fleischhauer00:2,Vasilev09,Gorshkov07,Yao05}.

Based on our recent discussion and demonstration of the single-photon generation process in the time domain \cite{Vasilev10,Nisbet11}, we here devise an analytical method for finding the control pulse required for the complete absorption of single photons of arbitrary temporal shape, i.e. for any given running-wave probability amplitude $\phi_{in}(t)$ impinging on a cavity mirror \cite{probability}. We emphasise that this technique applies directly to the implementation of a quantum memory and is pertinent to a variety of cavity-based systems. As will be discussed later, this scheme relates most obviously to mapping Fock-state encoded qubits to atomic states \cite{Cirac97,BBMEJ07}, but also very simply extends to other possible superposition states, e.g. photonic time bin or polarisation encoded qubits \cite{Wilk07-Science,Specht11,Sun04}. 

\begin{figure}
\centering\includegraphics[width=0.95\columnwidth]{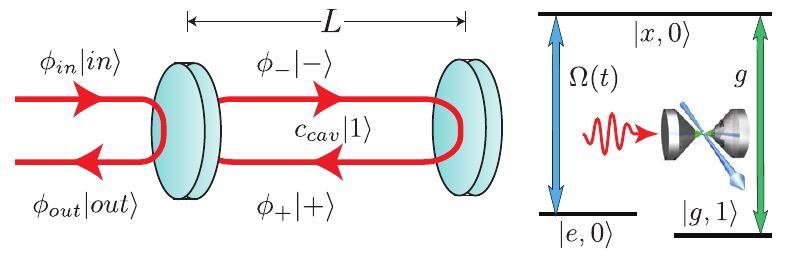}
\caption{(Color online) Coupling schemes. The left schematic shows the input-output cavity coupling formulation, whilst the right diagram shows the atomic states $|e\rangle$, $|x\rangle$ and $|g\rangle$ with their respective cavity photon number states $|0\rangle$ and $|1\rangle$. The couplings of the control pulse and cavity are given by $\Omega(t)$ and $g$ respectively. }
\label{levels}
\end{figure}

In the original transmission protocol \cite{Cirac97}, the leading and trailing edges of the photon change role upon its absorption by the second atom. The claim had therefore been made that such an approach would require photons of symmetric temporal shape; it has become evident that this is not a strict condition, and models have been made using a frequency-mode decomposition of free space to illustrate this mathematically \cite{Fleischhauer00:2,Lukin00}. Our procedure operates in the time domain and thus does not require any such decomposition into frequency modes. It is therefore closely analogous to the photon generation process, where the temporal shape of an emitted photon is unambiguously linked to the shape of the driving laser pulse. 

Prior to investigating the effect of the atom-cavity and atom-laser coupling, we shall briefly formulate a description of the input-output coupling of an optical cavity in the time domain. Inside the cavity, we assume that the mode spacing is so large that only one single frequency mode is involved, with the dimension-less probability amplitude $c_{cav}(t)$ for occupying the one-photon Fock state $|1\rangle$ at resonance frequency $\omega_{cav}$. Furthermore, we assume that one cavity mirror has a reflectivity of $100\,\%$, thus ensuring that coupling to the outside field is controlled uniquely by the field reflection and transmission coefficients, $r$ and $\tau$, of the other mirror. It is then very convenient to decompose the cavity mode into submodes $|+\rangle$ and $|-\rangle$, travelling towards and away from this mirror, respectively. In turn, the spatio-temporal representation of the cavity field reads
\begin{equation}
\phi_+(t)|+\rangle + \phi_-(t)|-\rangle,
\label{inout}
\end{equation}
with $\phi_-(t)=\phi_+(t)+\Delta\phi$, where any change of the running-wave probability amplitude from $\phi_+$ to $\phi_-$ at the mirror is taken into account with $\Delta\phi$.  The latter is small for mirrors of high reflectivity, such that $c_{cav}(t)\simeq \phi_+(t) \sqrt{t_r}\simeq \phi_-(t) \sqrt{t_r}$, and $t_r=2L/c$ the cavity's round-trip time (note that all running-wave probability amplitudes $\phi_{any}$ are of dimension $s^{-1/2}$, whereas the occupational probability amplitudes $c_{any}$ are dimensionless).  To properly describe the field outside the cavity, we decompose it also into incoming and outgoing spatio-temporal field modes, with running-wave probability amplitudes $\phi_{in}(t+z/c)$ and $\phi_{out}(t-z/c)$ for finding the photon in the $|in\rangle$ and $|out\rangle$ states at time $t$ and position $z$, respectively. The coupling mirror at $z=0$ acts as a beam splitter with the operator $a^\dagger_- (r a_+ + \tau a_{in}) + a^\dagger_{out} (\tau a_+ - r a_{in})$ coupling the four running-wave modes inside and outside the cavity. In matrix form, this coupling equation reads
\begin{equation}
\left(\begin{array}{c}\phi_++\Delta\phi \\ \phi_{out}\end{array}\right) 
=
\left(\begin{array}{c}\phi_-\\ \phi_{out}\end{array}\right) 
=
 \left(\begin{array}{cc}r & \tau \\ \tau & -r \end{array}\right) \left(\begin{array}{c}\phi_+\\ \phi_{in}\end{array}\right)
\label{an_cr}
\end{equation}
To relate $c_{cav}(t)$ to the running-wave probability amplitudes, we take $r\approx 1-\kappa t_r$ and $\tau=\sqrt{2\kappa t_r}$, where $\kappa$ is the polarisation decay rate of the cavity. Furthermore, we make use of
\begin{equation}
\frac{d c_{cav}}{dt} \simeq\frac{c_{cav}(t+t_r)-c_{cav}(t)}{t_r}= \frac{(\phi_--\phi_+)\sqrt{t_r}}{t_r} = \frac{\Delta\phi}{\sqrt{t_r}}.
\end{equation}
With these relations, the first line of Eq.\,(\ref{an_cr}) can be written as $c_{cav}/t_r + \dot{c}_{cav} = (1-\kappa t_r) c_{cav}/t_r +\sqrt{2\kappa}\phi_{in}$. Therefore Eq.\,(\ref{an_cr}) takes the form of a differential equation
\begin{equation}
\left(
\begin{array}{c}\dot{c}_{cav}\\ \phi_{out}\end{array}\right) = \left(\begin{array}{cc}-\kappa & \sqrt{2\kappa}\\ \sqrt{2 \kappa} & -r \end{array}\right) \left(\begin{array}{c}c_{cav}\\ \phi_{in}\end{array}\right)
\label{inputoutput}
\end{equation}
which describes the coupling of a resonant photon into and out of the cavity mode. We emphasise that any deviation from resonance with the cavity mode can be included in time-dependent phase factors of the probability amplitudes. Therefore the above coupling model applies to any case where just one cavity-field mode is involved, using the frequency $\omega_c$ of this mode as a carrier. We would like to emphasise that the result obtained from our simplified input-output model is fully equivalent to the conclusions drawn from the more sophisticated standard approach that involve a decomposition of the continuum into a large number of frequency modes \cite{Walls-Milburn}. No such decomposition is required here as we model the system uniquely in the time domain. Note also that the expansion of the present model to both mirrors being transparent is straight forward, but not required for the following analysis.

Next, we need to examine the coupling of a single atom to the field mode of the cavity. This has been discussed in a recent review \cite{Kuhn10} and the references mentioned therein. To best illustrate the process, we consider a three level $\Lambda$-type atom with two different electronically stable ground states $|e\rangle$ and $|g\rangle$, coupled by either the cavity field mode or the control laser field to one-and-the-same electronically excited state $|x\rangle$. For the one-photon multiplet of the generalised Jaynes-Cummings ladder, the cavity-mediated coupling between $|g,1\rangle$ and $|x,0\rangle$ is given by the atom-cavity coupling strength $g$, while the control laser  couples $|e,0\rangle$ with $|x,0\rangle$ with Rabi frequency $\Omega(t)$. The probability amplitudes of these particular three product states read $c_e(t)$, $c_g(t)$, and $c_x(t)$, respectively, with their time evolution  given by the master equation of the coupled atom-cavity system
\begin{equation}
\setlength\arraycolsep{1.4pt}
\left(
\begin{array}{c}\dot{c}_e \\ \dot{c}_x \\ \dot{c}_g \\ \phi_{out}\end{array}\right)=
 \left(\begin{array}{cccc}
0 & -i \Omega(t)^*/2 & 0 & 0\\
-i \Omega(t)/2 & -\gamma & -i g & 0\\ 
0 & -i g^* & -\kappa & \sqrt{2\kappa} \\ 
0&0 & \sqrt{2\kappa} & -r \end{array}
\right) \left(\begin{array}{c}c_e\\ c_x\\ c_g\\ \phi_{in}\end{array}\right),
\label{TDSE}
\end{equation}
which combines the Schr\"{o}dinger equation for a three-level atom coupled to the cavity with the input-output relation from Eq.\,(\ref{inputoutput}). Here  $\gamma$ is the spontaneous-emission rate of the atom in state $x$.  The coupling to the external field is taken into account by the mirror-induced  decay rate $\kappa$. 

Note that $c_g(t) \equiv c_{cav}(t)$  because the state $|g,1\rangle$ is the only atom-field product state in which there is one photon in the cavity.  We now restrict our discussion to the resonant case with the overall phase chosen such that $g$ and $\phi_{in}(t)$ are both real for all $t$. By consequence, the probability amplitudes $c_g(t)$ and $c_e(t)$ are real as well, while  $c_x(t)$ is purely imaginary. We  also emphasize that we are considering only one quantum of excitation, therefore at any given time, the probability of occupying $|g,0\rangle$ is given by the total probability of having a photon outside the cavity, i.e. either in state $|in\rangle$  or in $|out\rangle$.  The only way of coupling these states is via the cavity mirror to $|g,1\rangle$, which is taken into account by Eq.\,(\ref{TDSE}). 

Our goal is to completely absorb an incoming photon, with its wavepacket given by $\phi_{in}(t)$. Obviously, complete impedance matching is a necessary condition, i.e. $\phi_{out}(t)=0$ must be met at all times. We furthermore assume that $r\approx 1$, 
i.e. the reflectivity of the cavity mirror is nearly one.
Using these constraints with eqn. (\ref{TDSE}) we express two of the probability amplitudes in terms of the specified photon function $\phi_{in}(t)$ and its time derivative,
\begin{equation}
c_g(t) =\phi_{in}(t) / \sqrt{2\kappa} \label{cgg},
\end{equation}
\begin{eqnarray}
c_x(t) &=	&i  \left[ \dot{c}_g(t)-\kappa  c_g(t) \right]  /g^* \nonumber \\
		&=& i  \left[ \dot{\phi}_{in}(t)-\kappa  \phi_{in}(t) \right]  /g^*\sqrt{2\kappa}. 
\label{cxx}
\end{eqnarray}
We also obtain the product
\begin{equation}
\Omega(t)c_e(t)=
2    \left[ i \dot{c}_x(t)+ i \gamma  c_x(t) -g c_g(t) \right]\equiv\zeta(t). 
\label{OCE}
\end{equation}
To proceed, we need to consider the initial population of states and the overall continuity of probabilities (by which we mean the excitation is conserved). It is natural to assume that the atom-cavity system is initially prepared in state $|g,0\rangle$, which lies outside the considered subspace. The photon is completely in the incoming state $|in\rangle$, i.e. $\int|\phi_{in}(t)|^2dt = 1$. The portion that couples into the cavity therefore directly populates $|g,1\rangle$, so that the continuity balance yields
\begin{equation}
\rho_{ee}(t)=\rho_0-\rho_{gg}(t)-\rho_{xx}(t)+\int_{-\infty}^t[|\phi_{in}(t')|^2 - 2\gamma \rho_{xx}(t') ]dt'.
\label{REE}
\end{equation}
Here, $\rho_{ii}=c_i^*c_i$ are the populations and therefore the diagonal elements of the density matrix, and $2\gamma\rho_{xx}$ is the population decay rate from the electronically excited state. To account for an imperfect state preparation with  a small initial population in state $|e,0\rangle$, the offset term $\rho_0$ has been introduced phenomenologically. The relevance of this term (which is ideally zero) becomes obvious in the following discussion.

Equation (\ref{REE}) now gives a direct analytical expression for $\rho_{ee}$, and as $c_e$ is real on resonance, we simply divide Eq.\,(\ref{OCE}) by $\sqrt{\rho_{ee}(t)}$ to obtain
\begin{equation}
\Omega(t)=\frac{\zeta(t)}{\sqrt{\rho_{ee}(t)}}
=\frac{2 \left[ i \dot{c}_x(t)+ i \gamma  c_x(t) -g c_g(t) \right]}{\sqrt{\rho_{ee}(t)}}.
\label{OMT}
\end{equation}
This is an analytical expression, derived following equations (\ref{cgg}) to (\ref{REE}), for the Rabi frequency $\Omega(t)$ required to achieve full impedance matching over all times and therefore to absorb the incoming photonic wave packet $\phi_{in}(t)$ completely by the atom-cavity system. Most computer algebra systems (CAS) can be used to obtain a closed expression for $\Omega(t)$ given the a functional expression for the incoming photon. Of course, this non-iterative algorithm may also be applied numerically.
 
We now consider physically realistic photons that are normally restricted to a finite support of well-defined start and end times, $t_{start}$ and $t_{stop}$. We also assume they start off smoothly, i.e. with $\phi_{in}(t_{start})=\frac{d}{dt}{\phi}_{in}(t_{start})=0$, as described in \cite{Vasilev10}. However, the second derivative might be non-zero at $t_{start}$, so that Eqn. (\ref{OCE}) yields $\Omega(t_{start})c_e(t_{start})\neq 0$. To satisfy the latter inequality a small but non-vanishing initial population is required in the state $|e,0\rangle$. In other words, perfect impedance matching with $\rho_0 = 0$ would only be possible with photons of a physically impossible infinite duration. 

\begin{figure}
\centering\includegraphics[width=0.95\columnwidth]{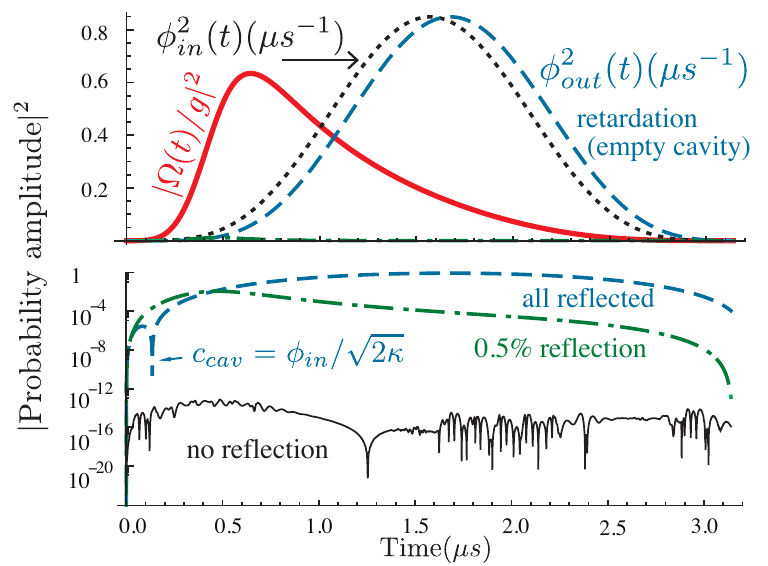}
\caption{(Color online) Incoming sin$^{2}$ photon (dotted). Case (a): Empty cavity, all reflected (dashed); Case (b): System prepared in $|g,0\rangle$, small reflection (dash-dotted); Case (c): Small initial population in $|e,0\rangle$, reflection suppressed (thin solid). The control pulse (thick solid) is derived to match case (c).}
\label{sin2}
\end{figure}

\begin{figure}[t]
\centering\includegraphics[width=0.95\columnwidth]{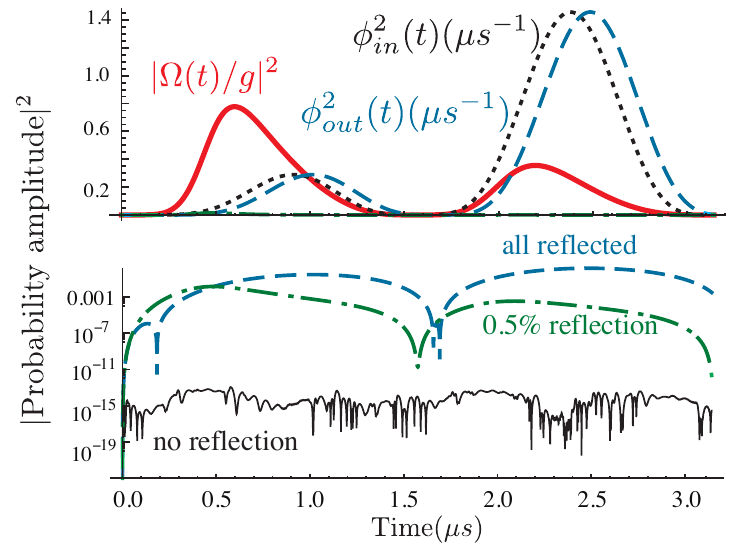}
\caption{(Color online) A twin peak pulse. All traces as described in fig.\,\ref{sin2}.}
\label{twins}
\end{figure}

To illustrate the power of the procedure and the implications of the constraints to the initial population, we now apply the scheme to a couple of typical photon shapes that one may obtain from  atom-cavity systems. To this purpose, we consider a cavity with parameters similar to one of our own experimental implementations, with $(g,\kappa,\gamma)=2\pi\times (15, 3, 3)\,$MHz and a resonator length of $L=100\,\mu$m.
As a first example, we assume that a symmetric photon with $\phi_{in}(t)\propto \sin^2(\pi t/\tau_{photon})$ impinges on the cavity. For a photon duration of $\tau_{photon}=3.14\,\mu$s, fig.\,\ref{sin2} shows $\phi_{in}(t)$,  $\Omega(t)$  and the probability amplitude of the reflected photon, $\phi_{out}(t)$, as a function of time. The latter is obtained from a numerical solution of eqn.\,(\ref{TDSE}) for the three cases of (a) an empty cavity, (b) an atom coupled to the cavity initially prepared in $|g,0\rangle$, with $\rho_0=0$, and (c) a small fraction of the atomic population initially in state $|e,0\rangle$, with $\rho_0=0.5\%$. In all three cases, the Rabi frequency $\Omega(t)$ of the control pulse is identical. It has been derived analytically assuming a small value of $\rho_0=0.5\%$ (this choice is arbitrary and only limited by practical considerations, as will be discussed later).  From these simulations, it is obvious the photon gets fully reflected if no atom is present (case a), albeit with a slight retardation due to the finite cavity build-up time. Because the direct reflection of the coupling mirror is in phase with the incoming photon and the light from the cavity coupled through that mirror is out-of phase by $\pi$, the phase of the reflected photon flips around as soon as $c_{cav}(t)=\phi_{in}(t)/\sqrt{2\kappa}$. This shows up in the logarithmic plot as a sharp kink in $\phi_{out}(t)$ around $t=0.13\,\mu$s.

The situation changes dramatically if there is an atom coupled to the cavity mode. For instance, with the initial population matching the starting conditions used to derive $\Omega(t)$, i.e. case (c) with $\rho_0=0.5\%$, no photon is reflected. The amplitude of $|\phi_{out}(t)|^2$ remains below $10^{-12}$, which corresponds to zero within the numerical precision. However, for the more realistic case (b) of the atom-cavity system well prepared in $|g,0\rangle$,  the same control pulse is not as efficient, and the photon is reflected off the cavity with an overall probability of $0.5\%$. This matches the ``defect'' in the initial state preparation, and can be explained by the finite cavity build-up time leading to an impedance mismatch in the onset of the pulse.  

We emphasise that this seemingly small deficiency in the photon absorption might become significant with photons of much shorter duration. For instance, in the extreme case of a photon duration $\tau_{photon}<\kappa^{-1}$, building up the field in the cavity to counterbalance the direct reflection by means of destructive interference is achieved most rapidly without any atom. Any atom in the cavity will act as a sink, removing intra-cavity photons. With an atom present, a possible alternative is to start off with a very strong initial Rabi frequency of the control pulse. This will project the atom-cavity system initially into a dark state, so that the atom does not deplete the cavity mode. Nonetheless, the initial reflection losses would still be as high as for an empty cavity.  

Second, we consider a more sophisticated non-symmetric twin-peaked photon impinging onto the cavity, with $\phi_{in}(t)\propto \sin^2(2 \pi t/\tau_{photon}) \cos(\pi/2 (1-t/\tau_{photon}))$. As in the first example, we simulate the process with a control-pulse Rabi frequency $\Omega(t)$ derived for $\rho_0=0.5\%$. From fig.\,\ref{twins}, it is obvious the results are the same as before, despite the rather complicated photon shape.  This impressively demonstrates the potential of our method, giving one a very simple procedure to derive control-laser pulses for absorbing incoming photons of arbitrary shape. 

\begin{figure}[t]
\centering\includegraphics[width=0.95\columnwidth]{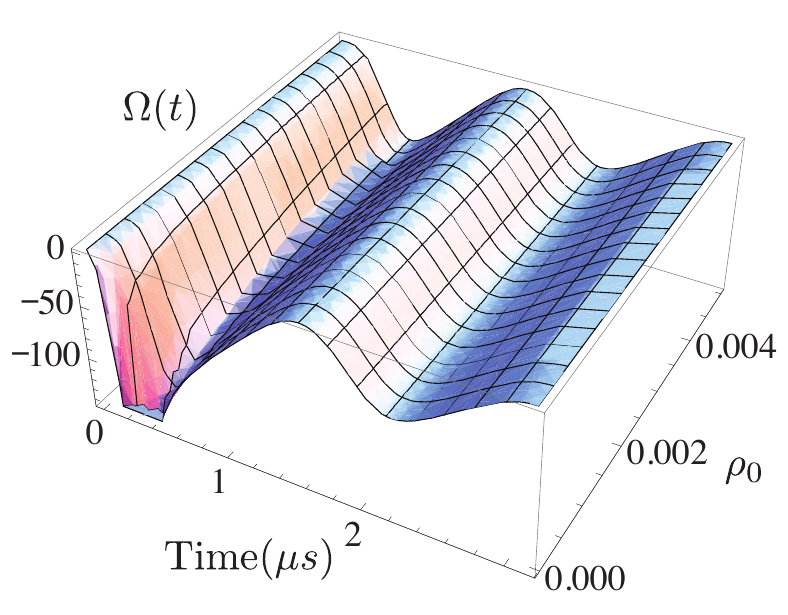}
\caption{(Color online) Derived Rabi frequency, $\Omega(t)$, as a function of $t$ and $\rho_0$. In this particular case, the required pulse-shape changes little for values of $\rho_0$ greater than 0.002. }
\label{limit}
\end{figure}

Finally, we investigate how $\Omega(t)$ changes in the limit $\rho_0\longrightarrow 0$, which would normally correspond to the initial condition in any implementation. For the twin-peak photons discussed above, fig.\,\ref{limit} shows $\Omega(t)$ for different values of $\rho_0$. As one would expect from our above discussion of the build-up dynamics, $\Omega(t)$ does not converge to a finite function in the limit $\rho_0\longrightarrow 0$. Hence for any realistic scenario, one needs to chose $\rho_0$ to be as small as possible such that  a feasible driving pulse shape is obtained. Limits to this are normally the finite laser power and/or the finite bandwidth of the amplitude modulators. For the `best possible' driving pulse derived in this manner, a numerical simulation starting with the atom-cavity system prepared in $|g,0\rangle$ then reveals the actual efficiency. In the particular examples discussed here, the photon-reflection losses are comparable to the non-zero $\rho_0$ used to derive the driving pulse, i.e. at a level below 0.5\% which is negligible compared to the noise affecting current experimental approaches. 

\begin{figure}[h]
\centering\includegraphics[width=0.95\columnwidth]{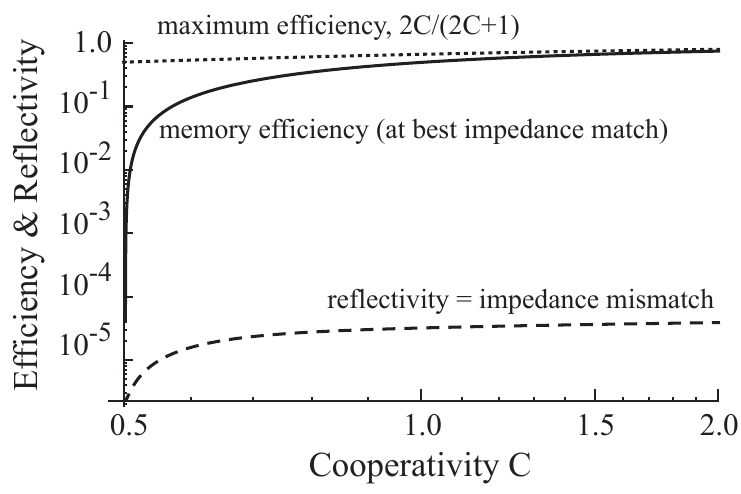}
\caption{Storage efficiency (probability of mapping the incoming photon to $|e,0\rangle$, solid line) and impedance mismatch (probability of back-reflecting the incoming photon, dashed line) as a function of the cooperativity, $C=g^2/(2\kappa\gamma)$. The dotted line shows the maximum possible efficiency. Both efficiency and impedance mismatch have been numerically calculated for $\kappa=\gamma=2\pi 3\,$MHz using symmetric $\sin^2$ pulses of $3.14\,\mu$s duration, with control pulses modelled according to Eqs.\.(\ref{cxx}-\ref{OMT}) to achieve best impedance matching.}
\label{lossy}
\end{figure}
A unique feature of the method presented here is its universal applicability to a wide range of atom-cavity coupling regimes.  The simple recipe for calculating the control pulse  assures full impedance matching for any given single-photon wavepacket impinging on the cavity. The present work therefore extends the recent approach made by Gorshkov et al.\,\cite{Gorshkov07}, as we need no \`{a} priori restriction to extreme coupling cases in their respective approximations. While the authors of the former were seeking for the best possible quantum memory, we here ask for perfect impedance matching. The latter is a necessary condition (satisfied by one and only one control pulse), which may not ensure 100\% efficiency under all circumstances. To illustrate the interplay of impedance matching and memory efficiency, Fig.\,\ref{lossy} shows the reflection probability and the memory efficiency (excitation transfer to $|e,0\rangle$) as a function of  the cooperativity, $C=g^2/(2\kappa\gamma)$. Obviously, the impedance matching condition is always met, but the efficiency varies. For $C>1$, it asymptotically reaches the predicted optimum \cite{Gorshkov07} of $2C/(2C+1)$, but it drops to zero at $C=1/2$ (i.e. for $g=\kappa=\gamma$). In this particular case, the spontaneous emission loss via the atom equals the transmission of the coupling mirror. Hence the coupled atom-cavity system behaves like a balanced Fabry-Perot cavity, with one real mirror being the input coupler, and the spontaneously emitting atom acting as output coupler. Therefore the photon gets not mapped to $|e,0\rangle$. This limiting case furthermore implies that impedance matching is not possible for $C<1/2$, as the spontaneous emission via the atom would then outweigh the transmission of the coupling mirror. The application of our formalism therefore fails in this weak coupling regime  (actually, the evaluation of Eq.\,(\ref{REE}) would then yield values of  $\rho_{ee}<0$, which is not possible).

\begin{figure}
\centering\includegraphics[width=0.95\columnwidth]{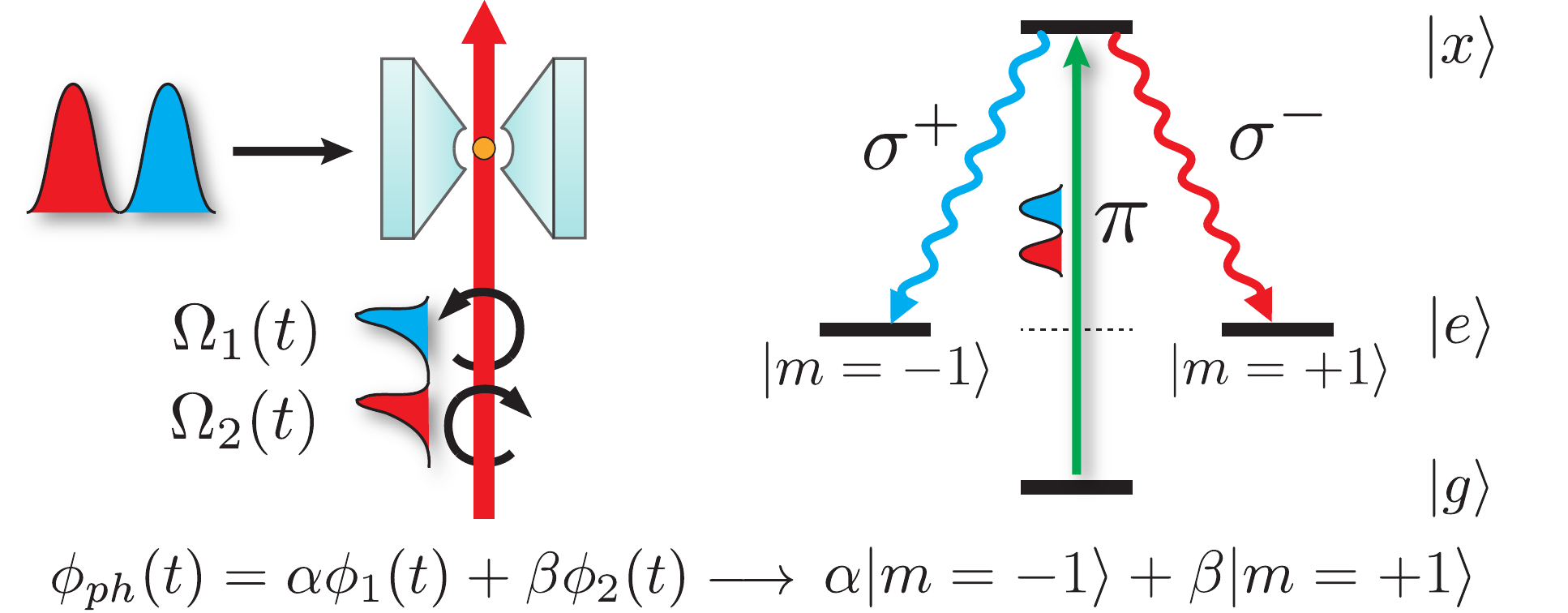}
\caption{(Color online) Mapping of a photonic time-bin superposition state to a superposition of atomic (spin) states. The left diagram shows a photonic qubit impinging on an atom-cavity system, and the respective driving pulses required in order to map the qubit to the atomic states (right).}
\label{mapping}
\end{figure}

\begin{figure}
\centering\includegraphics[width=0.95\columnwidth]{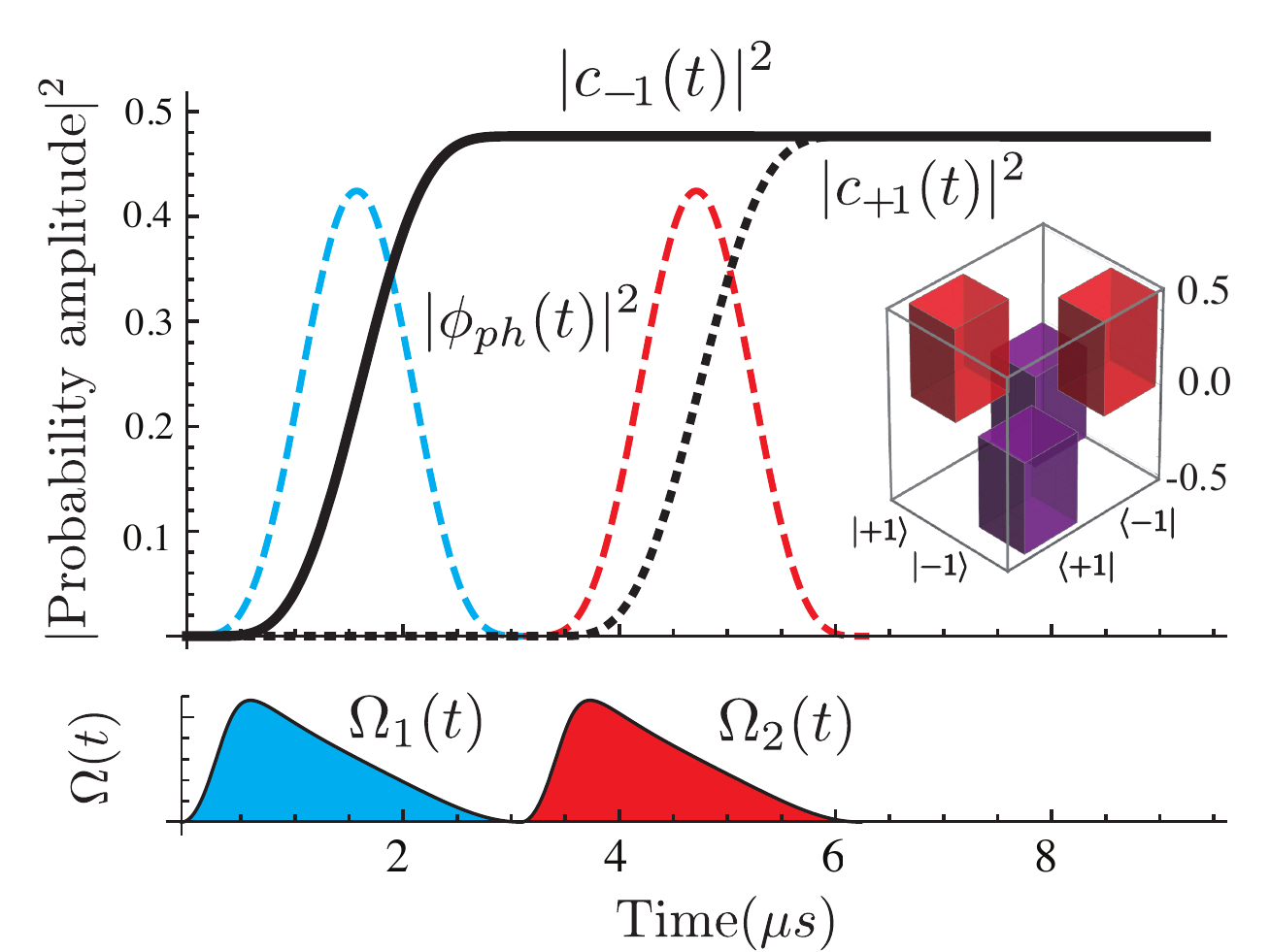}
\caption{(Color online) Simulation of the photon to atom state-mapping utilising the procedure outlined above. The photonic qubit is prepared in the state  $\frac{1}{\sqrt{2}}(\phi_1-\phi_2)$ and mapped onto the atomic spin states $|m=-1\rangle$ and $|m=+1\rangle$. The populations of the atomic states are shown (thick line:  $|m=-1\rangle$, thick dotted line: $|m=+1\rangle$), as is the incoming photonic state (thin dashed lines). The two driving pulses are shown below (arb. units), whilst the upper inset shows the atomic state density matrix after the successful absorption of the photon. }
\label{mapping2}
\end{figure}

We now discuss the application of our scheme to mapping an arbitrary time-bin superposition photonic state \cite{Marcikic02},  $\phi_{ph}(t)=\alpha\phi_{1}(t)+\beta\phi_{2}(t)$ to a superposition of atomic spins, $\alpha|m=-1\rangle+\beta|m=+1\rangle$. Assuming a $\pi$ polarised incoming photon and  $\phi_{1}$ and $\phi_{2}$  not overlapping in time, (i.e. they represent two orthogonal field modes), we devise two driving pulses $\Omega_{1}(t)$ and $\Omega_{2}(t)$ that map these two field modes into $|m=-1\rangle$ and $|m=+1\rangle$, respectively. As illustrated in fig. \ref{mapping}, the latter can be achieved with $\Omega_{1}(t)$ and $\Omega_{2}(t)$ having different circular polarisations. As a simple consequence of the unitarity of the time evolution, successive application of the two driving pulses, each within its respective time bin, results in the desired mapping of photonic to atomic superposition states. We emphasise that the pair of driving pulses only depends on the mode functions $\phi_{1,2}$ describing the two time bins, and not on their relative amplitudes $\alpha$ and $\beta$. Further, the three-level systems addressed during the first and second time bin are different from one another, and no cross talk occurs due to the different polarisations of the driving pulses. Hence the numerical simulation depicted in fig.\,\ref{sin2} now applies to the two time bins individually, with a specific example of the scheme shown in fig.\,\ref{mapping2}. The plot is calculated using the same values of $(g,\kappa,\gamma)$ as before, and demonstrates a storage efficiency of 95.3\% with a fidelity of unity. The latter is a consequence of the fact that the only incoherent process in the model, non-adiabatic excitation and decay to and from the excited state $|x,0\rangle$, is considered as a loss from the system. Thus it reduces the efficiency of the mapping, but does not contribute to any incoherence in the final atomic state superposition. 

One could also map a superposition of photonic number states to a pair of atomic states \cite{Cirac97,BBMEJ07}. For instance, $\alpha|0\rangle+\beta|1\rangle$ gets trivially mapped to $\alpha|g\rangle+\beta|e\rangle$; if no photon arrives then the atom remains unaffected (as the atom-cavity system stays in $|g,0\rangle$), whilst the signed driving pulse assures that a photon arriving in mode $\phi_{ph}(t)$ gets mapped to $|e\rangle$, in the way we depicted in fig.\,\ref{sin2}. Again, the unitarity of the time evolution assures the accurate mapping of superposition states. 

The photon reabsorption scheme discussed here, together with the earlier introduced method for generating tailored photons \cite{Vasilev10,Nisbet11}, constitute the key to analytically calculate  the optimal driving pulses needed to produce and absorb arbitrarily shaped single-photons (of finite support) with three level $\Lambda$-type atoms in optical cavities. This is a sine qua non condition for the successful implementation of a quantum network.  It is expected that this simple analytical method will have significant relevance for those striving to achieve atom-photon state transfer in C-QED experiments, where low losses and high fidelities are of paramount importance.

\acknowledgements
This work was supported by the Engineering and Physical Sciences Research Council\linebreak (EP/E023568/1), the Deutsche Forschungsgemeinschaft (Research Unit 635),  and the EU through the RTN EMALI (MRTN-CT-2006-035369).  We are grateful to G.\,S.\,Vasilev and D.\,Ljunggren for stimulating discussions in the early stages of this work.

\bibliographystyle{apsrev4-1}


\begin{thebibliography}{25}
\bibitem{probability}{$\phi_{in}(t)$ is the probability amplitude of the running photon, with $\int_{-\infty}^{+\infty}|\phi_{in}(t)|^2 dt =1$ and $|\phi_{in}(t)|^2dt$ the probability of the photon impinging on the mirror within $[t ... t+dt]$}\BibitemShut {NoStop}%
\end{thebibliography}

%
\end{document}